\documentclass[journal=mamobx,manuscript=article]{achemso}

\usepackage[version=3]{mhchem} 
\usepackage[T1]{fontenc}       
\usepackage{siunitx}
\usepackage{epstopdf}
\usepackage{amssymb}

\graphicspath{{graphics/}}

\renewcommand{\vec}[1]{\boldsymbol{#1}}
\newcommand*{\citen}[1]{%
  \begingroup
    \romannumeral-`\x 
    \setcitestyle{numbers}%
    \cite{#1}%
  \endgroup   
}

\author{Maud Formanek}
\affiliation{Centro de F\'isica de Materiales (CSIC, UPV/EHU) and Materials Physics Center MPC, Paseo Manuel de Lardizabal 5, E-20018 San Sebasti\'an, Spain}

\author{Angel J. Moreno}
\affiliation{Centro de F\'isica de Materiales (CSIC, UPV/EHU) and Materials Physics Center MPC, Paseo Manuel de Lardizabal 5, E-20018 San Sebasti\'an, Spain}
\alsoaffiliation{Donostia International Physics Center (DIPC), Paseo Manuel de Lardizabal 4, E-20018 San Sebasti\'an, Spain}

\email{angeljose.moreno@ehu.eus}
\phone{+34 943 01 8845}

\title[SCNPs under shear]
  {Single-Chain Nanoparticles \\ under Homogeneous Shear Flow}

\abbreviations{SCNP, MPCD, MD, VMD}

\begin{document}


\begin{abstract}
Single-chain nanoparticles (SCNPs) are a new class of macromolecular objects, synthesized through purely
intramolecular cross-linking of single polymer chains.
We use a multiscale hydrodynamics simulation approach to study, for the first time, SCNPs under shear flow.
We investigate the case of irreversible
SCNPs (permanent cross-links) in dilute solution. SCNPs emerge as a novel class of macromolecular objects with response to shear distinct from other systems such as linear chains, star polymers, rings or dendrimers.  This is evidenced by the observed  set of scaling exponents for the shear rate dependence of the SCNP static and dynamic properties. Surprisingly these exponents are, at most, marginally dependent on the specific topology of the SCNP (globular or sparse), suggesting that they are inherently related to the network-like character of the molecular architecture and not to its specific connectivity. At high Weissenberg numbers the dynamics of the sparse SCNPs is dominated by tumbling motion, whereas tank-treading predominates for the most globular SCNPs.

\end{abstract}

\section{I. Introduction}

Understanding the flow properties of macromolecular objects in solution is a problem of broad interest due to its relevance in many areas of soft matter, engineering and biophysics as e.g., microfludics, extrusion or blood flow. There is nowadays strong evidence that the non-equilibrium conformations and reorientational dynamics under shear flow have a strong, and sometimes even dramatic, dependence on other features than the macromolecular concentration and shear rate. Thus, polymers under shear flow have been shown to exhibit a rich variety of dynamic behaviors depending on the type of bonding potentials\cite{Sablic2017a, Chen2013},  excluded volume interactions \cite{Chen2015a}, hydrodynamics \cite{Ripoll2006, Sablic2017a, Liebetreu2018} and, specially, on the molecular architecture\cite{Winkler2014,Chen2017, Jeong2017,Jaramillo-Cano2018a}. The two most commonly observed reorientational behaviors at high Weissenberg numbers (i.e., when the characteristic time of the flow is shorter than longest molecular relaxation time) are: (a) tumbling motion, which is characterized by the polymer alternatingly adapting stretched and collapsed conformations over the course of which it flips `head' over `tail'  and (b) tank-treading  motion, during which the overall shape of the polymer stays approximately constant and aligned with the flow, while the individual monomers perform a rotation around the center-of-mass. Flexible linear chains are the archetypical example of polymers performing tumbling motion, as extensively discussed theoretically \cite{Puliafito2005, Winkler2006}, computationally\cite{Doyle1997, Hur2000, Schroeder2005, Dalal2012, Lang2014} and experimentally \cite{Smith1999, Hur2001, Teixeira2005, Gerashchenko2006}. Tank treading has been found in weakly deformable soft objects as vesicles and fluid droplets\cite{Skotheim2007, Abkarian2007, Dodson2010, Yazdani2011}. For polymers of more complicated architectures, such as stars or rings, these two motions are not only hard to define (for example, what are the `ends' of a ring to determine tumbling?), but also difficult to distinguish \cite{Chen2013, Chen2015, Sablic2017a}. 
Another remarkable effect of the molecular architecture is the observation of different sets of exponents for the shear rate dependence
of the dynamic observables (viscosity, rotational frequency, etc) as well as the static observables probing aligment and deformations along the relevant directions (flow, gradient and vorticity) \cite{Chen2017}.

In this article we present detailed results that constitute, to the best of our knowledge, the first investigation on 
the static and dynamic properties under shear flow of the so-called single chain nanoparticles (SCNPs) \cite{Altintas2016,lyon2015,reviewcsr,mavila2016,hanlon2016,PomposoSCNPbook}.
These topologically complex soft nano-objects, which are obtained through purely intramolecular cross-linking of functionalized single linear chains, are the basis of the so-called single-chain technology, a rapidly growing research area of enormous potential for use as biosensors \cite{gillisen2012single}, catalysts \cite{terashima2011,perez2013endowing,huerta2013,tooley2015}, 
drug delivery vehicles \cite{hamilton2009,sanchez2013design}, rheological agents \cite{Mackay2003,Arbe2016,Bacova2017}, etc. Inspired by biomacromolecules such as proteins or enzymes, it is a long-term goal \cite{PomposoSCNPbook} to develop SCNPs via intramolecular collapse/folding with accurate control of the sequence and architecture, and with high-performance and quick response (due to their internal malleability) to environmental changes (pH, temperature, stress, etc). 
A series of works combining small angle neutron and X-ray scattering (SANS and SAXS) with computer simulations \cite{sanchez2013design,Basasoro2016,Arbe2016,GonzalezBurgos2018} have revealed that
the standard protocols of synthesis in good solvent conditions produce topologically polydisperse SCNPs, 
with a distribution dominated by sparse arquitectures. 
The fundamental physical origin of this observation is that, independently of the specific chemical composition of the precursor \cite{Pomposo2014a,Pomposo2017rever}, the conformations of the SCNP in the good solvent conditions of synthesis are self-avoiding random walks (the polymer size scaling with the number of monomers as
$R \sim N^{\nu}$, where $\nu \approx 0.59$ is the Flory exponent \cite{Rubinstein2003}). In these configurations the formation of long-range loops is unfrequent, and most of the bonding events involve reactive groups separated by short contour distances, which are inefficient for the global compaction of the nanoparticle \cite{Moreno2013,LoVerso2014,LoVerso2015,Rabbel2017,Formanek2017,Oyarzun2018}.
It is worth mentioning that the SCNP structure has interesting analogies with intrinsically disordered proteins (IDPs)
\cite{Moreno2016JPCL,Moreno2018}, starting with their scaling behavior ($R \sim N^{\nu}$, $\nu \sim 0.5$). 
Though, unlike IDPs, SCNPs lack regions with ordered secondary structure, they still contain weakly deformable `domains' (dense clusters of permanents loops) connected by flexible segments. The peculiar architecture of SCNPs leads them 
to collapse  to a so-called fractal globular structure \cite{Grosberg1988,Mirny2011,Halverson2014} 
in crowded solutions and melts, suggesting this as a potential scenario for the effect of the steric crowding on IDPs in cell environments \cite{Moreno2016JPCL,GonzalezBurgos2018}.

Given their architectural complexity, their potential for a broad set of applications in solution and in bulk, and the mentioned structural analogies with IDPs, SCNPs are appealing systems for the investigation of physical properties in flow. 
The first simulations of SCNPs under (homogeneous) shear flow presented here are limitted to the case of high dilution. Properties in crowded solutions will
be explored in future work. We have made use of the 
multi-particle collision dynamics \cite{Malevanets1999, Malevanets2000} technique, which correctly implements hydrodynamic interactions on long time scales.
Several remarkable features are already found at high dilution. Thus, SCNPs emerge as a novel class of soft objects with a response to shear distinct from other flexible macromolecules such as linear chains, star polymers, rings or dendrimers.  The differences with these architectures manifest in 
the set of characteristic exponents found for the shear rate dependence of the SCNP static and dynamic properties. 
Unexpectedly, these exponents show no significant dependence on the specific architecture of the SCNP, in spite of the broad distribution of
investigated SCNP topologies (from globular to sparse ones).
This suggests that the observed exponents are inherently related to the network-like character of the molecular architecture, but not to the specific connectivity of the network. At high Weissenberg numbers the dynamics of the sparsest SCNPs is dominated by tumbling motion, whereas tank-treading motion is 
predominant for the most globular ones. 

The article is organized as follows. In Section II we give model and simulation details. Structural and dynamic observables under shear flow are characterized and discussed in Section III. Conclusions are given in Section IV.

\section{II. Model and simulation details}
We use a multi-scale hybrid simulation technique that combines molecular dynamics (MD) for the polymers with multi-particle collision dynamics (MPCD) for the solvent. The precursors of the SCNPs are modeled as linear chains of $N=200$ monomers, of which a fraction $f = N_r/N = 0.25$ are functional reactive monomers. These have the ability to form irreversible crosslinks and are distributed randomly across the polymer backbone, with the constraint that they are never placed consecutively in order to avoid trivial crosslinks. 
We employ the coarse-grained Kremer-Grest bead-spring model \cite{Kremer1990} to simulate both the precursor molecules and the synthesized SCNPs. As such, the non-bonded interactions between any two given monomers are modeled by a purely repulsive Lennard-Jones (LJ) potential,
\begin{equation}
U^{\rm LJ}(r) = 4\epsilon \left[ \left(\frac{\sigma}{r}\right)^{12} -\left(\frac{\sigma}{r}\right)^{6} +\frac{1}{4}\right] \, ,
\end{equation}
to account for excluded-volume interactions. Here, $r = \vert \vec{r}_i - \vec{r}_j\vert$ is the euclidean distance between monomers $i$ and $j$, while $\epsilon/k_{\rm B}T = 1$ and $\sigma = 1$ set the units of energy and length, respectively. We use a cutoff distance $r_{\rm c} = 2^{1/6}\sigma$, at which both the potential and the corresponding forces are continuous. In addition, bonded monomers along the contour of the chain and cross-linked monomers interact via a finitely extensible nonlinear elastic (FENE) potential \cite{Kremer1990}, 
\begin{equation}
U^{\rm FENE}(r) = - \epsilon K_{\rm F} R_0^2 \ln \left[ 1 - \left(  \frac{r}{R_0}\right)^2 \right] \, , 
\label{eq:fene}
\end{equation}
with $K_{\rm F} = 15\sigma^{-2}$ and $R_0 = 1.5\sigma$. This combination of LJ and FENE potentials limits the fluctuation of bonds and guarantees chain uncrossability, as well as mimicking good solvent conditions. 

Our simulation protocol consists of two steps. First, 50 SCNPs are synthesized under highly dilute conditions without hydrodynamic interactions by employing Langevin dynamics simulations \cite{Izaguirre2001}. The stochastic nature of the cross-linking process leads to all of the synthesized SCNPs having a distinct topology. A detailed description of the implementation of the cross-linking process can be found in Ref.~\citen{Moreno2013}. Briefly, permanent cross-links are formed between two reactive monomers that have not yet cross-linked to any other one and that are separated by less than the capture distance $r_{\rm b} = 1.3\sigma$. In case there are multiple possible cross-linking partners for any given monomer at any given time, one of them is chosen at random. Finally, after the formation of a bond, the two involved monomers interact \textit{via} the FENE potential introduced in  \ref{eq:fene} for the remainder of the simulation.   

\begin{figure}[ht]
\centering
  \includegraphics[height=7.5cm]{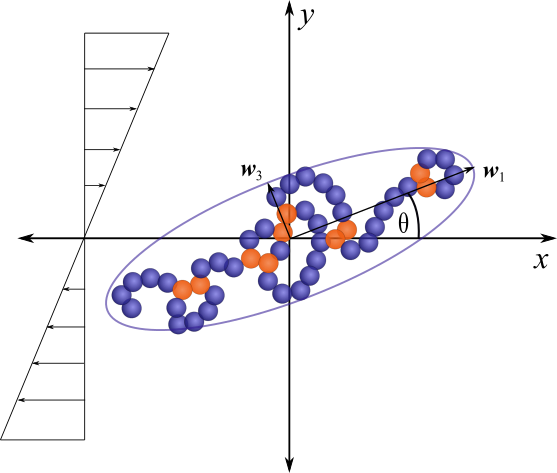}
  \caption{Schematic illustration of the simulation setup, indicating the fluid velocity profile as well as the eigenvalues  and eigenvectors of the gyration tensor. Here, $x$ is the flow direction, $y$ the gradient direction and $z$ -- pointing perpendicular to the plane -- is the vorticity direction. $\theta$ is the angle between the principal vector of the gyration tensor, $\vec{\bf w}_1$, and the flow direction. Reactive monomers forming cross-links are colored in orange, the rest in blue.}
  \label{fgr:scheme}
\end{figure}

Next, we sort the 50 synthesized SCNPs into 6 groups based on the value of their asphericity parameter (see \ref{eq:asph}) and choose one from each group at random for the simulations under shear flow. In this way we study the effect of the SCNP shape by taking representatives of the whole distribution
of asphericities.
The solvent is modelled \textit{via} MPCD \cite{Malevanets1999, Malevanets2000}, a mesoscopic particle-based technique for hydrodynamics simulations. 
The solvent is composed of $N_s$ point-like particles of mass $m$, which correspond to individual volumes of the fluid. The MPCD algorithm consists of two alternating steps which govern the dynamics of the solvent: i) A \textit{streaming} step,
in which the solvent particles are propagated according to ballistic motion for a time $h$:
\begin{equation}
\vec{\bf r}_i (t+\Delta t) = \vec{\bf r}_i(t) + h \vec{\bf v}_i(t) \, , 
\label{eq:streaming}
\end{equation} 
with $\vec{\bf r}_i$ and $\vec{\bf v}_i$ the position and velocity of the solvent particle $i$.
ii) A \textit{collision} step, in which they exchange linear momentum. To achieve this, the particles are sorted into cubic cells of length $a$ and subjected to a rotation around a random axis by an angle $\alpha$ with respect to the center-of-mass velocity of the cell $\vec{\bf v}_{\rm cm}$, i.e.
\begin{equation}
\vec{\bf v}_i (t + \Delta t) = \vec{\bf v}_{\rm cm}(t) + \mathbf{R}(\alpha)\left(\vec{\bf v}_i(t) - \vec{\bf v}_{\rm cm}(t)\right) \, ,
\label{eq:SR} 
\end{equation} 
with $\mathbf{R}(\alpha)$ the rotation matrix.
This conserves the total mass, linear momentum and energy of the system. To satisfy Galilean invariance, the cubic grid used to sort the particles has to be shifted randomly in each of the 3 directions at each collision step\cite{Ihle2001, Ihle2003}. A linear shear profile $\langle v_x(y) \rangle= \dot{\gamma} y $  is introduced \textit{via} Lees-Edwards boundary conditions\cite{Lees1972}. In the former expression $\dot{\gamma}$ is the shear rate, $v_x$ the component of the velocity in the flow direction
and $y$ the coordinate in the simulation box along the gradient direction. \ref{fgr:scheme} presents the setup of the simulation, showing the velocity profile and indicating the flow ($x$) and gradient ($y$) directions.  

Since shear flow leads to viscous heating, a cell-level Maxwell-Boltzmann scaling thermostat is employed to keep the fluid at a fixed temperature $T$. Finally, the monomers of the SCNPs are coupled to the solvent by including them in the collision step (\ref{eq:SR}). It should be noted that this coupling is strong enough to ensure thermalization of the polymers as well as the solvent. Between successive collisions, the SCNP is propagated in time according to Newton's equations of motion, which are integrated using the Velocity Verlet algorithm\cite{Frenkel1996} with a time-step of $\Delta t = 0.01$. The number of solvent particles per cell is $\rho = 5$, their mass is $m = 1$, while the mass of the solute monomers is $M = \rho m = 5$.  The remaining parameters are $\alpha =\SI{130}{\degree}$, $h = 0.1\sqrt{ma^2/k_BT}$ and $a = \sigma = 1$. Furthermore, the volume of the simulation box $V = L_xL_yL_z$ is chosen based on the radius of gyration $R_{\rm g}$ of the SCNP at equilibrium ($\dot{\gamma}=0$), such that $L_\mu = 50\sigma \geq 4R_{\rm g} $ for $\mu \in \lbrace y, z\rbrace$. The extension of the box in the $x$-direction is increased with higher shear rates to account for the stretching of the SCNPs in the flow direction and ranges from $L_x = 50\sigma$ to $L_x = 100\sigma$.  
We perform 20 independent simulation runs for each shear rate $\dot{\gamma},$ with different initial conformations and velocities for each of the 6 individual SCNP topologies. Each run consists of an equilibration period over $10^{5}$ MD steps and a production cycle of $10^7$ MD steps. 

For comparison with a linear reference system, we have performed analogous simulations and analysis for a linear chain with the same number of monomers,
$N=200$, as the SCNP. The corresponding observables discussed in the article for the SCNPs are shown in the Supporting Information for the linear chains
(Figs S1-S4). These complement data for linear chains in the literature, which have been reported for much shorter chains ($N \le 60$) \cite{Aust1999}, 
and for long DNA chains (combining experiments and numerical modelling) \cite{Schroeder2005}.
In spite of the differences in the used models (and in particular in the implementation of the 
interactions with the solvent), we find good agreement between the sets of exponents of the linear chains reported here and in those works (see \ref{tab:scaling}).

\section{III. Results and discussion}

\subsection{IIIA. Structural properties}
\begin{figure}[ht]
\centering
  \includegraphics[width=0.97\linewidth]{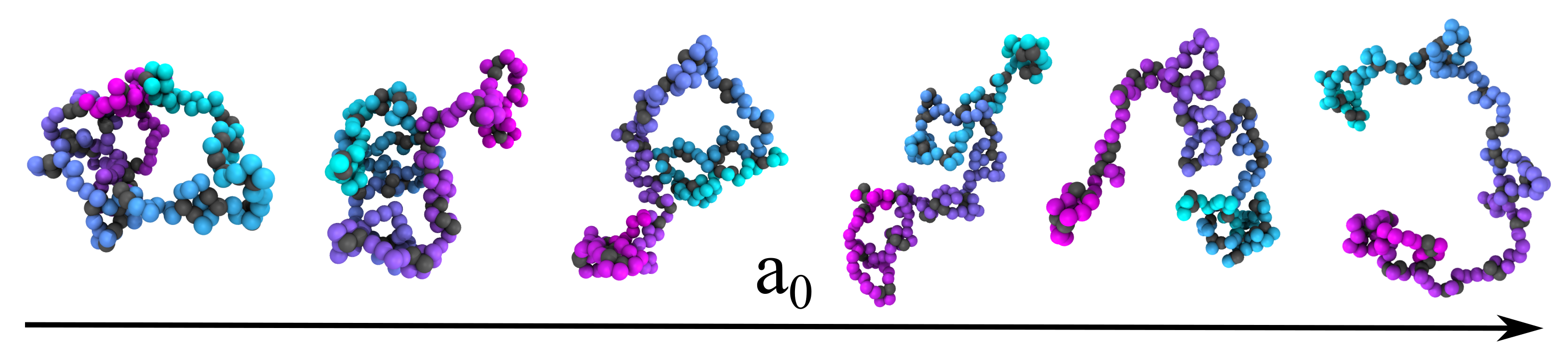}
  \caption{Representative snapshots of SCNPs at $\dot{\gamma}=0$
  with different values of the equilibrium asphericity $a_0$ (increasing from left to right).
  Grey beads are cross-linked monomers. The rest of the monomers are colored, from magenta to cyan, according to their position
  in the backbone of the linear precursor.}
  \label{fgr:snaps}
\end{figure}

The conformational properties of a polymer can be described by the gyration tensor
\begin{equation}
G_{\mu\nu} =\frac{1}{N} \sum_{i=1}^{N}  (r_{i,\mu}-r_{{\rm cm},\mu})(r_{i,\nu}-r_{{\rm cm},\nu}) \, , 
\end{equation}
where $r_{i,\mu}$ and $r_{{\rm cm},\mu}$ are the $\mu$-th Cartesian components of the position of monomer $i$ and the  center-of-mass, respectively. We calculate the eigenvalues $\lambda_1 \geq \lambda_2 \geq \lambda_3$ as well as the eigenvectors $\vec{w}_i$ of the gyration tensor, which define an ellipsoid with the same inertial properties as the polymer. Various shape descriptors can be derived from the components as well as from the eigenvalues of the gyration tensor, such as the asphericity,
\begin{equation}
a = \frac{(\lambda_2-\lambda_1)^2 + (\lambda_3-\lambda_1)^2 + (\lambda_3-\lambda_2)^2}{2(\lambda_1+\lambda_2+\lambda_3)^2} \, ,
\label{eq:asph}
\end{equation}
which ranges from 0 for objects with spherical symmetry to 1 for a one-dimensional object ($\lambda_2 = \lambda_3 = 0$). 
For each (topologically different) individual SCNP, its asphericity is averaged over its 20 independent trajectories.
The equilibrium asphericity $a_0 = a(\dot{\gamma}=0)$ can be used to discern SCNPs in terms of their structure and it has been shown to correlate with the relative deformability $\delta = \sqrt{\left(\langle R_{\rm g}^2\rangle - \langle R_{\rm g}\rangle^2\right)/\langle R_{\rm g}^2\rangle}$ of individual SCNPs\cite{Moreno2016JPCL,Moreno2018}. $R_{\rm g}$
is the molecular radius of gyration, 
\begin{equation}
R_{\rm g}^2 = \frac{1}{N}\sum_{i=1}^{N} ( \vec{\bf r}_i - \vec{\bf r}_{\rm cm})^2 \, .
\end{equation}
\ref{fgr:snaps} shows representative snapshots of 6 topologically different SCNPs in equilibrium ($\dot{\gamma}=0$), 
covering the whole range of $a_0$-values. The topological polydispersity in the SCNPs
ranges from globular objects (left) to very open ones similar to self-avoiding chains (right), though the topological
distribution is dominated by sparse architectures.

\begin{figure}[ht]
\centering
  \includegraphics[width=0.5\linewidth]{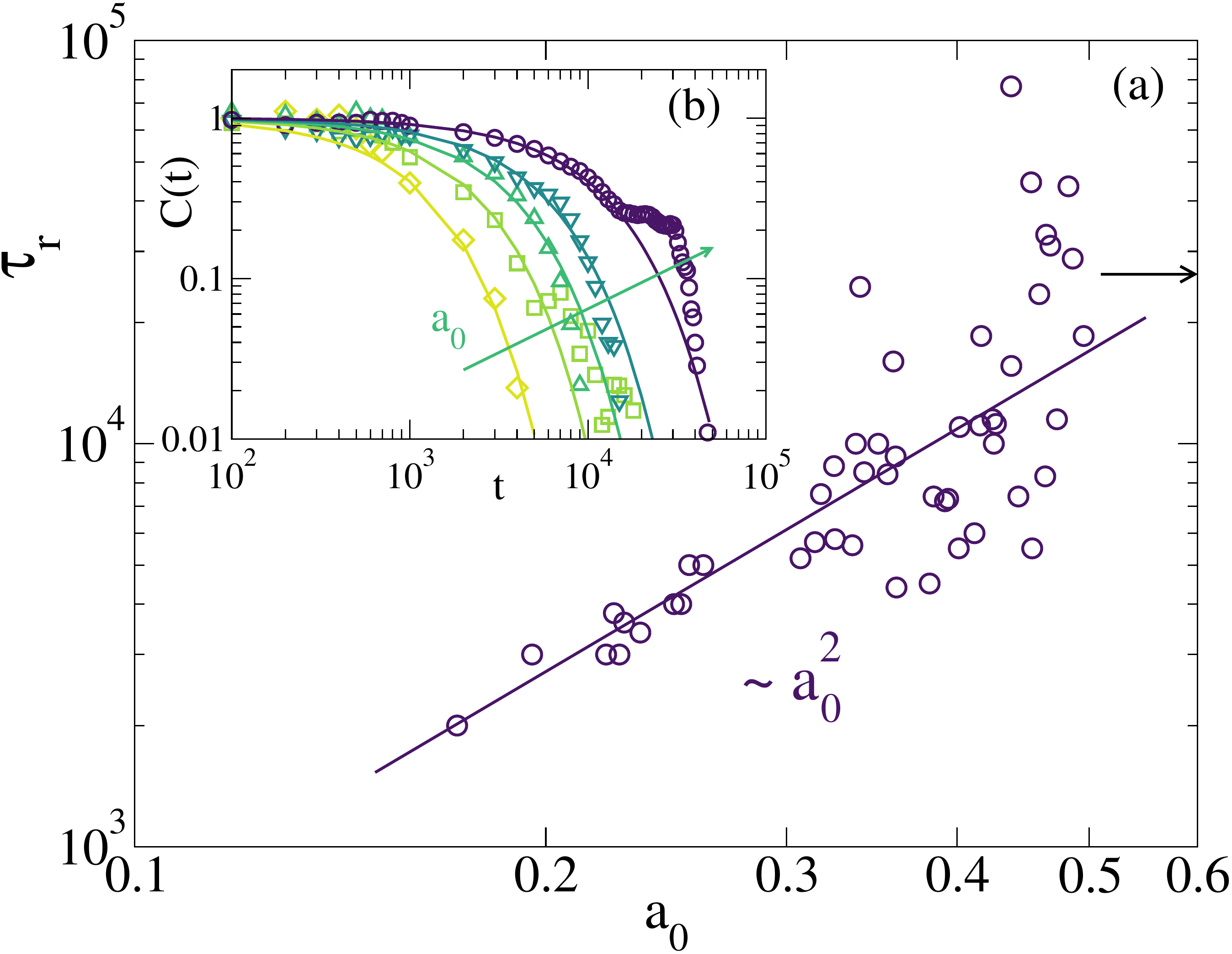}
  \caption{(a) Symbols: equilibrium ($\dot{\gamma}=0$) relaxation times $\tau_{\rm r}$ vs. equilibrium asphericity parameters $a_0$ of 50 topologically distinct SCNPs. The arrow indicates the $\tau_{\rm r}$ of the linear chain. The inset (b) shows the autocorrelation function $C(t)$ of the radius of gyration $R_{\rm g}$ used to determine the relaxation times for 5 typical SCNPs. The arrow indicates increasing $a_0$.}
  \label{fgr:asphericityrelaxation}
\end{figure}

Since the specific topology of the individual SCNP may be expected to influence its response to shear and its relevant time scales, we introduce the dimensionless Weissenberg number, $Wi = \dot{\gamma}\tau_{\rm r}$, to scale the shear rates. This parameter quantifies the ratio between the relaxation time $\tau_{\rm r}$ 
of the polymer at equilibrium and the characteristic time $\dot{\gamma}^{-1}$ of the shear flow. To determine the relaxation time of the individual SCNPs, we introduce the autocorrelation function of the radius of gyration at zero shear rate,
\begin{equation}
C(t) = \frac{\langle R_{\rm g}(t) R_{\rm g}(0)\rangle - \langle R_{\rm g}\rangle^2}{\langle R_{\rm g}^2\rangle- \langle R_{\rm g}\rangle^2}\, , 
\label{eq:autocorr}
\end{equation} 
and we define the relaxation time $\tau_{\rm r}$ as the time at which $C(t)$ has decayed to 0.2. The inset of  \ref{fgr:asphericityrelaxation} shows the autocorrelation function for representative SCNPs with very different topologies as measured by their deviation from spherical symmetry. We find that for most of the SCNP topologies $C(t)$ can be described by an exponential decay. 
\ref{fgr:asphericityrelaxation}a demonstrates that the equilibrium asphericity $a_0$ correlates with the relaxation time $\tau_{\rm r}$ obtained from $C(t)$. The times
encompass two decades and roughly scale as $\tau_{\rm r} \sim a_0^2$.

\begin{figure}[ht]
\centering
  \includegraphics[width=0.5\linewidth]{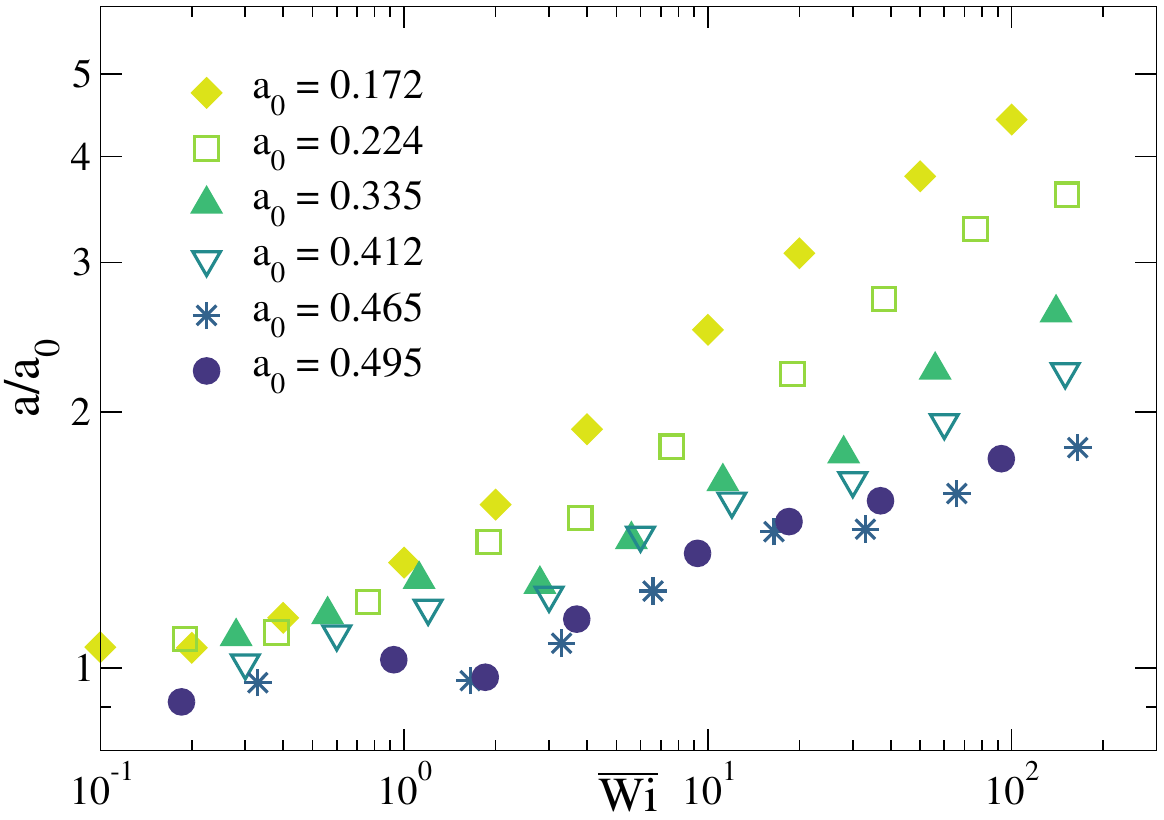}
  \caption{Normalized asphericity $a$ as a function of the Weissenberg number $Wi$.}
  \label{fgr:asphericity}
\end{figure}

\begin{figure}[ht]
\centering
  \includegraphics[width=1\linewidth]{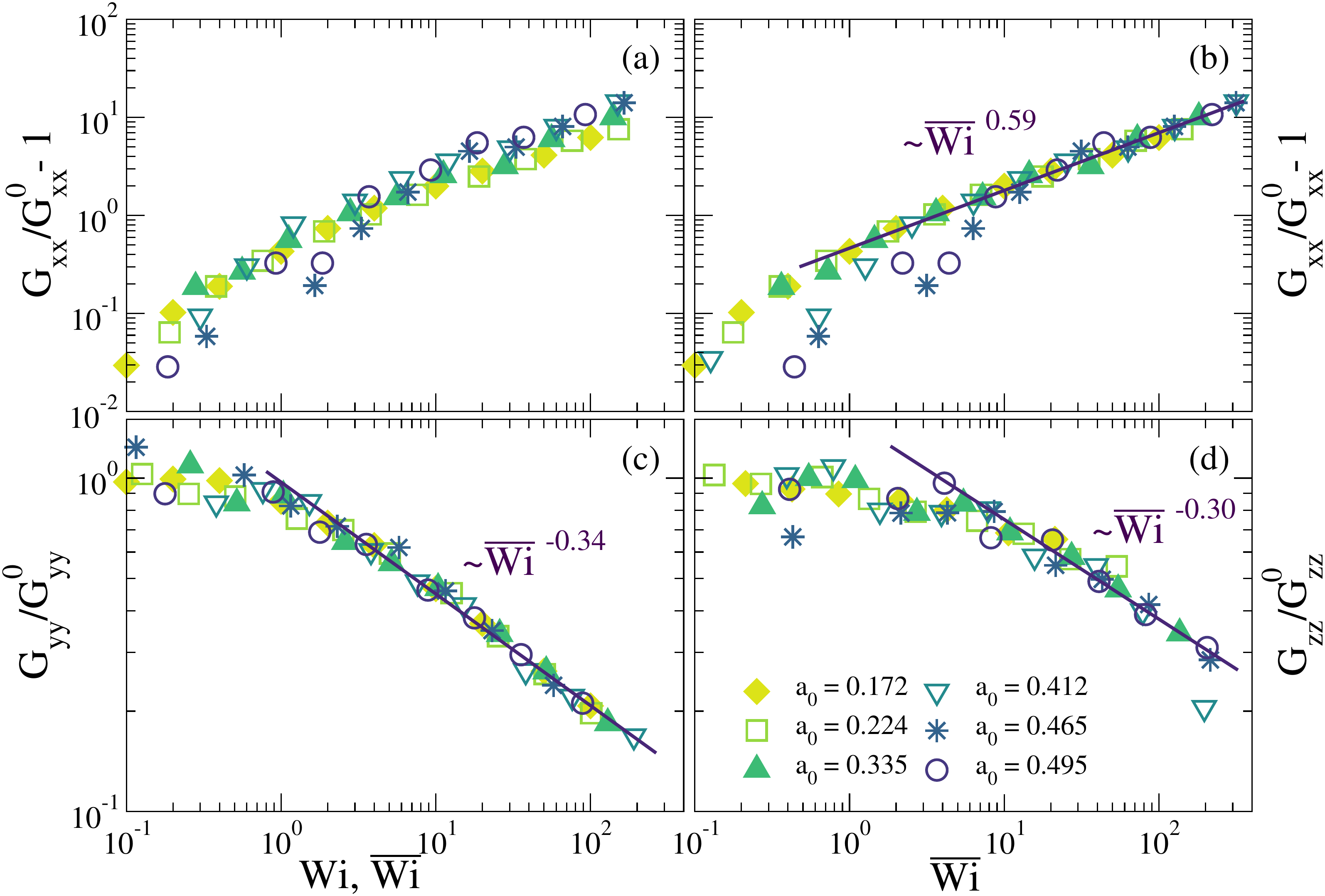}
  \caption{Normalized diagonal components of the gyration tensor $G_{\alpha\alpha}$ as a function of (a) Weissenberg number $Wi$ and (b-d) rescaled Weissenberg number $\overline{Wi}$ for various  SCNP topologies of distinct equilibrium asphericities. Results are given for: (a,b) flow direction, $G_{xx}$; (c) gradient direction, $G_{yy}$; (d) vorticity direction, $G_{zz}$. Symbol codes have the same meaning across all panels. Lines are fits to power laws.}
  \label{fgr:GxxGyy}
\end{figure}

To investigate the effect of the shear flow on the structure of the SCNPs, in  \ref{fgr:asphericity} we plot the normalized asphericity $a/a_0$ against the Weissenberg number $Wi$ for 6 different SCNPs of equilibrium asphericities  covering the whole range of $a_0$ values observed. Two regimes are clearly discernable: at small shear rates, where $Wi \ll 1$, the global shape of the SCNPs remains essentially undisturbed. When the longest relaxation time of the polymer $\tau_{\rm r}$  exceeds the characteristic time of the flow $\dot{\gamma}^{-1}$, i.e. in the region where $Wi \gg 1$, the SCNP reorients in the flow field and adopts less spherical conformations. The fact that the cross-over between the two response regimes falls in the region where $Wi \approx 1$ for all SCNP topologies confirms the consistency of our method of determining the relaxation time $\tau_{\rm r}$. SCNPs of different topologies display different responses under high shear rates, with the shapes of SCNPs of low equilibrium asphericities $a_0$ being more affected than those of high $a_0$ values
---still, it should be noted that actually the sparsest SCNPs ($a_0 \sim 0.5$) cannot be elongated beyond the rod limit ($a =1$) that is already approached at $Wi \sim 100$ (see \ref{fgr:asphericity}).

Additional insights into the structural changes induced by the shear flow can be gained from considering the diagonal components of the gyration tensor, $G_{\alpha\alpha}$.  \ref{fgr:GxxGyy} displays the  components in the flow ($G_{xx}$), gradient ($G_{yy}$) and vorticity ($G_{zz}$) directions normalized by their
values at zero shear rate ($G_{\alpha\alpha}^0$). Contrary to the changes in the asphericity with the shear rate, we find that the data for the different SCNPs collapse onto a master curve, when we rescale the Weissenberg number by a factor of the order of unity. This scaling factor should be inherently connected to the specific topology of the SCNP, in a similar fashion to, e.g., the case of star polymers, where the master curve is obtained after rescaling $Wi$ by a factor dependent on the number of arms (functionality $f$) \cite{Ripoll2006,Chen2017}.
Panels (a) and (b) of \ref{fgr:GxxGyy} depict the diagonal component $G_{xx}$ before and after rescaling the Weissenberg number and demonstrate that only a small scaling factor is required in order to obtain a master curve. For the remainder of the article, we will present the simulation results as a function of this rescaled Weissenberg number, $\overline{Wi}$.
In the regime of high shear rates, all the diagonal components of the gyration tensor follow  power laws,
with exponents depending on the direction (flow, gradient, vorticity) but, within statistics, independent of the specific SCNP topology. The SCNPs stretch in the flow direction,  $G_{xx} \sim Wi^{\mu}$ with $\mu = 0.59$, whereas they are compressed in both the gradient as well as the vorticity direction ($\mu = -0.34$ and -0.30, respectively). As it will be shown, master curves $X \sim \overline{Wi}^{\mu}$ are obtained for the rest of the observables $X$ computed for the SCNPs, with exponents depending on $X$  but not on the specific topology of the SCNP (characterized by the equilibrium asphericity $a_0$). This result is  unexpected given the very different topologies of SCNPs covering a rather broad range of asphericities (see \ref{fgr:snaps}), and therefore it seems related to the network-like topology of the SCNP but not the specific connectivity of the network. 

As can be seen in \ref{tab:scaling}, the stretching and compression behaviors probed by the components $G_{\alpha\alpha}$ of the gyration tensor are also
observed as power laws, with different exponents, in other molecular architectures. 
The relative stretching of the SCNPs along the flow direction is similar to that of linear chains and rings.
This is consistent with the, in average, sparse character of the SCNP topology. The weaker dependence on the shear rate found
for the SCNPs ($\mu = 0.59$ vs. 0.63 and 0.65 for linear chains and rings, respectively) likely originates from the
cross-linked character of the SCNP that hinders elongation with respect to linear chains and rings.
The apparently much stronger relative deformation of stars and dendrimers in the flow direction is 
analogous to the corresponding observation for the asphericity (\ref{fgr:asphericity}), i.e., molecular architectures that are
already sparse in equilibrium  have less margin of relative elongation  than the spherical but still highly malleable
stars and dendrimers.

The SCNPs show similar exponents for the relative compression along the gradient ($\mu = -0.34$) and vorticity ($\mu = -0.30$)
directions. Regarding the compression along the gradient direction the SCNPs show a weaker $Wi$-dependence than linear chains and rings,
and in this case SCNPs behave similar to dendrimers and high-$f$ stars. A tentative explanation for this feature might be that the presence
of cross-links (in the SCNPs) or branch points in the dense structures of dendrimers and high-$f$ stars, combined with jamming, hinders compression in comparison with linear chains and rings. However, the SCNPs stretch in the flow direction more similarly to linear chains and rings because they frequently have long outer ends or loops that facilitate orientation of the principal axes with the flow.
Regarding compression in the vorticity direction, the most malleable systems (linear chains, rings, low-$f$ stars and SCNPs)
show similar exponents $\mu \sim 0.3$, where a weak or marginal dependence on $Wi$ is found for the high-$f$ stars 
and dendrimers.

\begin{table}[h]
  \caption{Scaling exponents for the $Wi$-dependence (at $Wi > 1$) of different static and dynamic observables,
  normalized by the values at $\dot{\gamma}=0$, in SCNPs and
   other molecular architectures: linear chains, rings, 4th generation dendrimers (G4D) and star polymers.
   In the stars $f$ is the number of arms (functionality) and $N_{\rm a}$ the number
   of monomers per arm.  $N$ is the total number of monomers in the macromolecule ($N = fN_{\rm a}$ for the stars).
    Bold fonts: results from this work. Normal fonts: simulation results from the literature (references are indicated). Values in the third column were obtained by combining simulations and experiments in DNA chains \cite{Schroeder2005}.
   Values with star superscripts are not given in the original references; 
   we have obtained them by sampling and fitting the data reported there.}
  \label{tab:scaling}

\begin{tabular}{| p{2.1cm} | p{1.41cm} | p{1.25cm} | p{1.25cm} | p{1.49cm} |  p{1.43cm} |p{2.92cm} | p{1.41cm} | }
\hline
                         & Linear $N=200$   & Linear (DNA)                         & Linear $N \le 60$     & Ring  $N \le 120$                & G4D $N=62$                                 & Star \hspace{2cm} $2N_{\rm a} \le 80$                                                              & SCNP $N=200$  \vspace{3 mm}  \\
\hline
$G_{xx}/G^0_{xx}-1 $    & {\bf 0.63}       &                                      &                       &  0.65 \cite{Chen2015}            & 0.86 \cite{Nikoubashman2010}               &  1.0  ($f \le 50$) \cite{Ripoll2006} 0.90$^{*}$ ($f=18$)  \cite{Jaramillo-Cano2018}                & {\bf 0.59}  \vspace{3 mm}  \\
\hline
$G_{yy}/G^0_{yy}   $    & {\bf -0.48}      & -0.50 \cite{Schroeder2005}           &                       & -0.41 \cite{Chen2013,Chen2015}   & -0.30$^{*}$ \cite{Nikoubashman2010}        & -0.42 ($f \le 10$) \cite{Chen2017} \hspace{1cm} -0.32$^{*}$ ($f=18$) \cite{Jaramillo-Cano2018}     & {\bf -0.34}  \vspace{3 mm}  \\ 
\hline
$G_{zz}/G^0_{zz}   $    & {\bf -0.34}      & -0.34 \cite{Schroeder2005}           &                       & -0.32 \cite{Chen2013,Chen2015}   & $\approx 0^{*}$ \cite{Nikoubashman2010}    & -0.29 ($f \le 10$) \cite{Chen2017} \hspace{1cm}-0.14$^{*}$ ($f=18$) \cite{Jaramillo-Cano2018}      & {\bf -0.30}  \vspace{3 mm}  \\
\hline
$m_{\rm G} $             & {\bf 0.53}       & 0.57 \cite{Schroeder2005}            & 0.54 \cite{Aust1999}  &  0.60 \cite{Chen2013SM,Chen2013} & 0.49 \cite{Nikoubashman2010}               &  0.63 ($f \le 10$) \cite{Chen2017}  0.65 ($f \le 50$) \cite{Ripoll2006,Jaramillo-Cano2018}         & {\bf 0.67}  \vspace{3 mm}  \\
\hline
$\omega_z/\dot{\gamma}$  & {\bf -1.0}       &                                      &                       & -0.38 \cite{Chen2013SM,Chen2013} &                                            & -0.52 ($f \le 10$) \cite{Chen2017} \hspace{1cm} -1.0  ($f \le 50$) \cite{Ripoll2006}               & {\bf -0.75}  \vspace{3 mm}  \\
\hline
$\eta/\eta^0 $          & {\bf -0.66}      & -0.61 \cite{Schroeder2005}           & -0.59 \cite{Aust1999} & -0.43 \cite{Chen2013,Chen2015}   &                                            & -0.40 ($f \le 10$) \cite{Chen2017}                                                                 & {\bf -0.48}  \vspace{3 mm}  \\
\hline
$\Phi_1/\Phi_1^0 $      & {\bf-1.2}        & -1.37 \cite{Schroeder2005}            & -1.2 \cite{Aust1999}  & -0.97 \cite{Chen2015}            &                                            & -1.1 ($f \le 10$) \cite{Chen2017}                                                                  & {\bf -1.2}   \vspace{3 mm}  \\
\hline
\end{tabular}
\end{table}

\begin{figure}[ht]
\centering
  \includegraphics[width=0.5\linewidth]{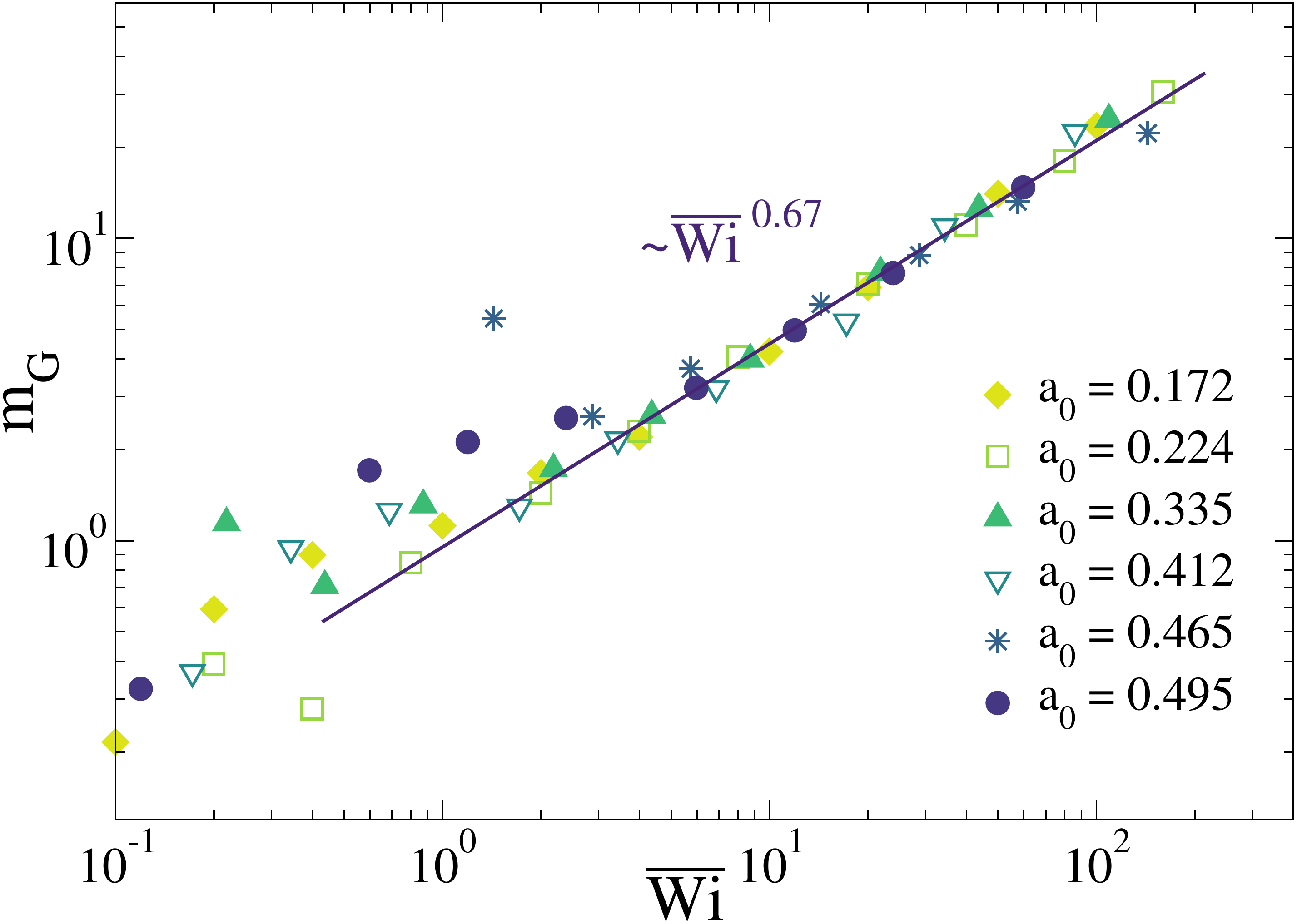}
  \caption{Orientational resistance $m_{\rm G}$ as a function of the rescaled Weissenberg number $\overline{Wi}$
  for SCNPs with different equilibrium asphericities.}
  \label{fgr:mg}
\end{figure}

At this point it should be noted that, though a broad range of $Wi$-values of experimental interest is investigated
here and in the references of \ref{tab:scaling}, the exponents given for the $G_{\alpha\alpha}$ components, and for the other
observables discussed in the rest of the article, may be just effective values in a slow crossover regime to the limit
of high Weissenberg numbers ($Wi \gg 100$). For example, scaling arguments predict for linear chains an intermediate regime
$G_{yy} \sim Wi^{-1/2}$ prior to the final regime $G_{yy} \sim Wi^{-2/3}$ 
reached in the high $Wi$-limit \cite{Winkler2010,Winkler2014}.

A measure of the average alignment of a macromolecule with the flow direction in the presence of shear is given by the orientational resistance parameter\cite{Bossart1995} 
\begin{equation}
m_{\rm G} = Wi\tan(2\theta) = Wi\frac{2G_{xy}}{G_{xx}-G_{yy}} \, ,
\label{eq:mg}
\end{equation}
which can be directly calculated from the gyration tensor. It is related to the angle $\theta$ between the eigenvector $\vec{{\bf w}}_1$ corresponding to the largest eigenvalue $\lambda_1$ and the flow direction $x$ (see  \ref{fgr:scheme}). For rodlike particles and linear polymers \cite{Doi1986,Link1993,Aust1999}
at low $\dot{\gamma}$ the components of the gyration tensor in \ref{eq:mg} scale as $G_{xy} \sim \dot{\gamma}$ and $\left(G_{xx}-G_{yy}\right) \sim \dot{\gamma}^2$, such that at low shear rate $m_{\rm G}$ becomes independent of $Wi$. Computational investigations employing the same MPCD method as in this study have confirmed this behaviour also for star polymers\cite{Ripoll2006}. 

As can be seen in  \ref{fgr:mg}, however, the data for SCNPs do not clearly approach a plateau at low $Wi$, though no conclusions can be made
due to the poor statistics at such low shear rates. Similar observations have been also found in, e.g., Ref.~\citen{Nikoubashman2010}.
The poor statistics originates from the denominator $G_{xx}-G_{yy}$ in \ref{eq:mg}, since at low shear rates the macromolecular
conformations along the flow and gradient directions are weakly perturbed with respect 
to the equilibrium values $G_{\rm xx}^0 = G_{\rm yy}^0$.
Moreover, the SCNPs simulated here are longer (larger $N$) than many of the previously studied systems (see \ref{tab:scaling}),
so that for the same value of the Weissenberg number the corresponding shear rate is even lower than in those systems. 
Still, we find that the data of $m_{\rm G}$ for different SCNP topologies again collapse onto a single curve upon applying the rescaling of the Weissenberg number. At high $Wi$, the orientational resistance follows a power law $m_{\rm G} \sim Wi^{\mu}$ with exponent $\mu = 0.67$, which is comparable to that found for star polymers, $\mu = 0.65$\cite{Ripoll2006, Jaramillo-Cano2018}, but considerably bigger than the one for linear chains, $\mu \sim 0.55$\cite{Aust1999, Schroeder2005}, or dendrimers, $\mu = 0.49$\cite{Nikoubashman2010}. While in star polymers strong resistance to alignment with the flow originates from jamming
in the two elongated bundles of arms (oriented in opposite directions in the stretched stars) \cite{Ripoll2006}, in the SCNPs it might be explained by the existence of permanent loops and clusters of loops along the SCNP. This results again in strong repulsive forces, stemming from jamming
within such loops and clusters, when the SCNP adopts stretched conformations.  

\subsection{IIIB. Rheological properties}
The contribution of the polymer to the viscosity of the dilute solution can be calculated from the Kramers-Kirkwood stress tensor\cite{Bird1987},
\begin{equation}
\sigma_{\mu\nu} = \sum_{i=1}^N \langle r_{i, \mu} F_{i, \nu}\rangle \, , 
\end{equation}
where $\vec{\bf F}_i$ is the total force acting on monomer $i$ and $\mu,\nu$ denote the Cartesian components. The polymer contribution to the viscosity is then given by
\begin{equation}
\eta = \frac{\sigma_{xy}}{Wi} \, .
\end{equation}  

\begin{figure}[ht]
\centering
  \includegraphics[width=0.5\linewidth]{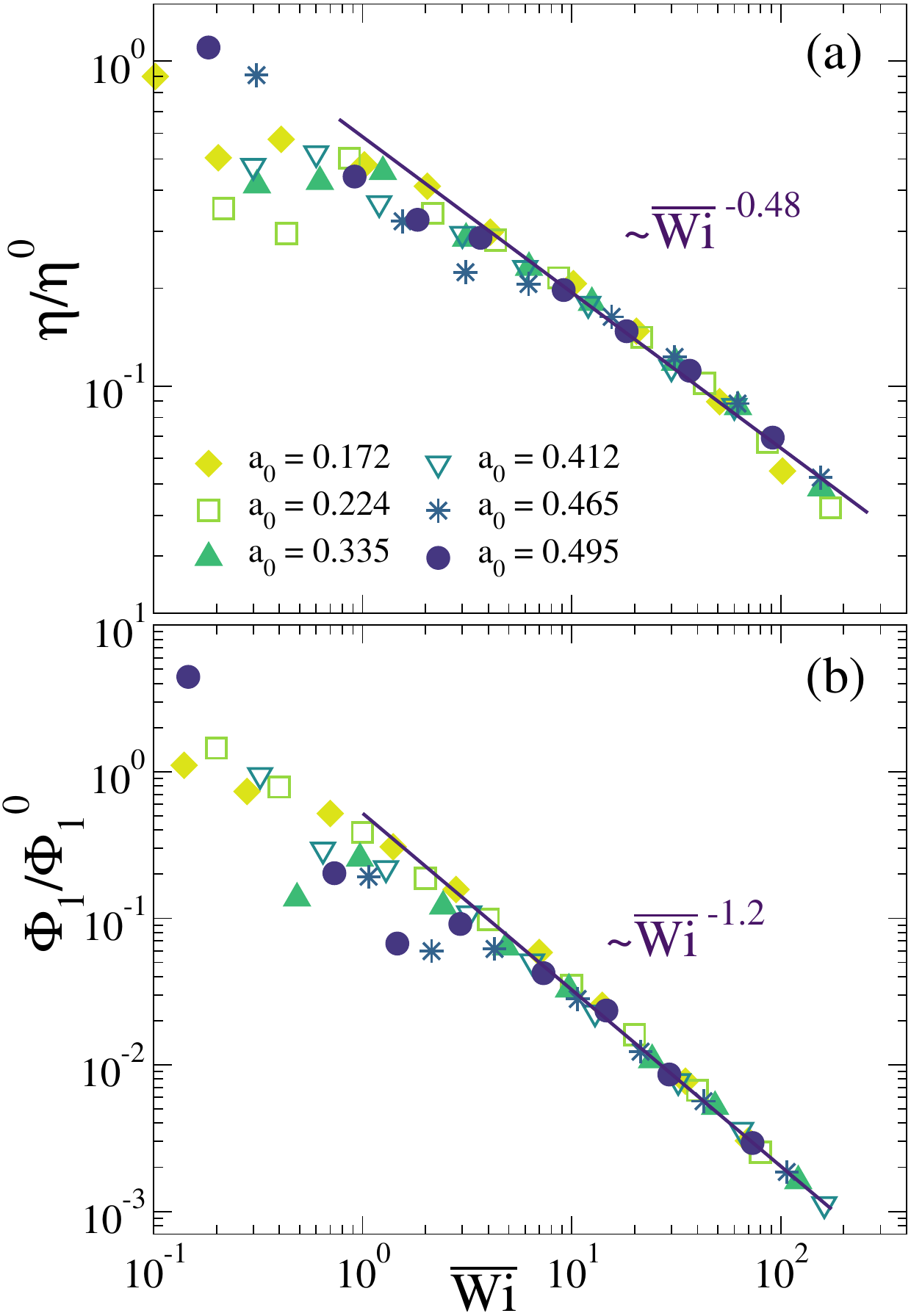}
  \caption{Normalized (a) viscosity $\eta$ and (b) first normal stress coefficient $\Phi_1$ as a function of the rescaled Weissenberg number $\overline{Wi}$ for SCNPs with different equilibrium asphericities.}
  \label{fgr:viscosity}
\end{figure}

\ref{fgr:viscosity}a shows the viscosity $\eta$ relative to the zero shear viscosity $\eta_0$. Here $\eta$ should be understood as
the intrinsic viscosity \cite{Larsonbook}, since data correspond to the limit of dilute SCNPs. The definition of $\eta$ includes a division by $Wi$.
Therefore $\eta_0$ is usually determined by the Newtonian plateau of $\eta$ at low shear rates $Wi \ll 1$. However, not all the SCNPs exhibit such a well-defined plateau, which is likely due to the big statistical uncertainty in $\sigma_{\mu\nu}$ at low Weissenberg numbers. 
Therefore, we normalize the viscosities such that the data collapse onto a single curve at high Weissenberg numbers. We find that the intrinsic viscosity of the SCNPs decreases with increasing Weissenberg number, alluding to shear-thinning behaviour, and that it scales as $\eta/\eta_0 \sim Wi^{\mu} $
with $\mu = -0.48$. This exponent is similar to those of stars and rings (see \ref{tab:scaling}) and much smaller than the value found for linear chains ($\mu \sim -0.6$; $\mu = -2/3$ in the limit of high Weissenberg numbers \cite{Winkler2010,Winkler2014}). The stronger $Wi$-dependence of the viscosity in the linear chains can be rationalized by the lower compactness of their self-avoiding structure with respect to the other systems, and hence the higher 
concentration of solvent around each monomer.

In the absence of hydrodynamic interactions, the stress tensor can also be calculated according 
to the Giesekus approximation \cite{Larsonbook}, which leads to
\begin{equation}
\eta = \sum_{i=1}^N \left\langle \frac{r_{i, y} r_{i, y}}{2}\right\rangle \, , 
\end{equation}
and thus $\eta \sim G_{yy}$. Agreement between $\eta$ and $G_{yy}$
has been found for linear chains in the free-draining limit\cite{Doyle1997, Schroeder2005}, 
and for semidilute solutions of linear chains with hydrodynamic interactions \cite{Huang2010} (where these are effectively screened out
by the concentration). In the case of dilute macromolecules with hydrodynamic interactions the relation $\eta \sim G_{yy}$
clearly breaks for linear chains and SCNPs, as evidenced by the rather different 
exponents for the $Wi$-dependence of $\eta$ and $G_{yy}$ (see \ref{tab:scaling}). This reflects the relevance of the
hydrodynamic interactions in the rheological properties of linear chains and SCNPs, apparentely playing a stronger role
than in low-$f$ stars or rings, where $\eta$ and $G_{yy}$ show very similar scaling with the Weissenberg number (see \ref{tab:scaling}).

In addition to the viscosity, we can also calculate the first normal stress coefficient from the stress tensor, i.e.,
\begin{equation}
\Phi_1 = \frac{\sigma_{xx}-\sigma_{yy}}{Wi^2} \, . 
\end{equation}
\ref{fgr:viscosity}b shows the first normal stress coefficient as a function of the rescaled Weissenberg number $\overline{Wi}$. 
Since $\sigma_{xx}-\sigma_{yy} \sim \dot{\gamma}^2$, a plateau is expected at low Weissenberg numbers.
However, similar to the viscosity data, poor statistics at $Wi < 1$ complicates the identification of the zero-shear value in terms of a plateau, 
so we again scale $\Phi_1$ such that the data collapse onto a master curve at high $Wi$. We find that the first normal stress coefficient decreases as $\Phi_1 \sim Wi^{\mu}$, with $\mu = -1.2$. This reflects a slightly weaker dependence than in linear 
polymers ($\mu \sim -1.3$; $\mu = -4/3$ according to scaling arguments) \cite{Aust1999,Huang2010, Winkler2014}, and slightly stronger than in 
low-$f$ stars and rings ($\mu = -1.1$ and -0.97)\cite{Chen2015}.   

\subsection{IIIC. Dynamic behavior}

\begin{figure}[ht]
\centering
  \includegraphics[width=0.5\linewidth]{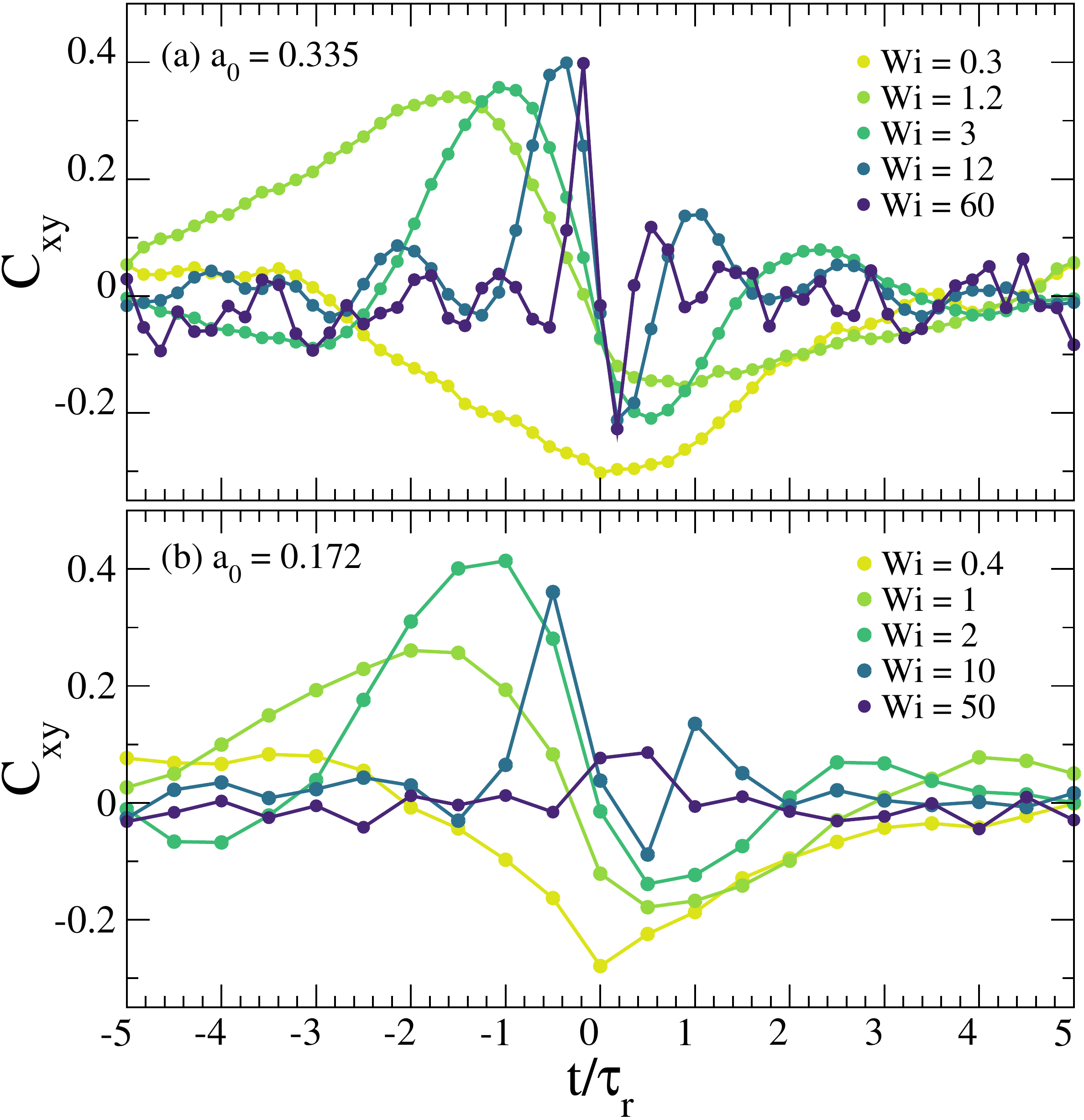}
  \caption{Flow-gradient extensional cross-correlation function $C_{xy}(t)$ for a SCNP of (a) intermediate and (b) low equilibrium asphericity. Times are rescaled by the longest relaxation time of the SCNP, $\tau_{\rm r}$.}
  \label{fgr:Cxy}
\end{figure}

Tumbling motion is typically analyzed \textit{via} the cross-correlation of the diagonal elements of the gyration tensor in the flow and gradient directions\cite{Teixeira2005, Huang2011, Chen2013, Chen2015} 
\begin{equation}
C_{xy}(t) = \frac{\langle \delta G_{xx}(0) \delta G_{yy}(t) \rangle}{\sqrt{\langle\delta G_{xx}^2(0)\rangle \langle \delta G_{yy}^2(0)\rangle}} \, , 
\end{equation}
where $\delta G_{\alpha\beta} = G_{\alpha\beta} - \langle G_{\alpha\beta}\rangle$ is the fluctuation of a component of $G$ around its mean value. Why this is a useful measure for detecting tumbling can be understood by looking at what happens during one tumbling event: for the majority of the time, the polymer is expanded in the flow direction to reduce stress from the current. However, thermal fluctuations lead to stochastic extensions of parts of the polymer in the gradient direction, whereupon those monomers experience an increased drag force from the flow and are pulled along the flow direction. As a result, the polymer as a whole contracts to a coil, flips around and subsequently extends again in the flow direction, with the `head' and the `tail' having switched sides. Thus, tumbling is characterized by negative anti-correlation peaks in the cross-correlation function $C_{xy}$. Despite $C_{xy}$ not being perfectly periodic but decaying to zero after some time, signatures of tumbling motion are clearly seen in \ref{fgr:Cxy}.

Tank-treading, quite contrary to tumbling, is characterized by a lack of fluctuations of the polymer extension in either direction, while the individual monomers rotate around the polymer center-of-mass, leaving the overall shape unaffected. We follow the approach used by Chen \textit{et al.}\cite{Chen2013} to detect tank-treading behavior in ring polymers by calculating the angular auto-correlation function: 
\begin{equation}
C_{\rm angle}(t) = \frac{\langle A(0)A(t)\rangle}{\langle A^2(0)\rangle}
\label{eq:cangle}
\end{equation}
where $A(t) = \sin\left(2\beta\right)$ and $\beta$ represents the angle between the vector connecting an individual monomer to the center-of-mass and the instantaneous first principal component of the polymer configuration. The averages in \ref{eq:cangle} are performed over the $N$ monomers of the SCNP.
If a polymer undergoes tank-treading, 
$C_{\rm angle}$ is expected to show an oscillating signal that is damped in time due to decorrelations arising from
the intramolecular motions in the deformable polymer configuration.

\begin{figure}[ht]
\centering
  \includegraphics[width=0.5\linewidth]{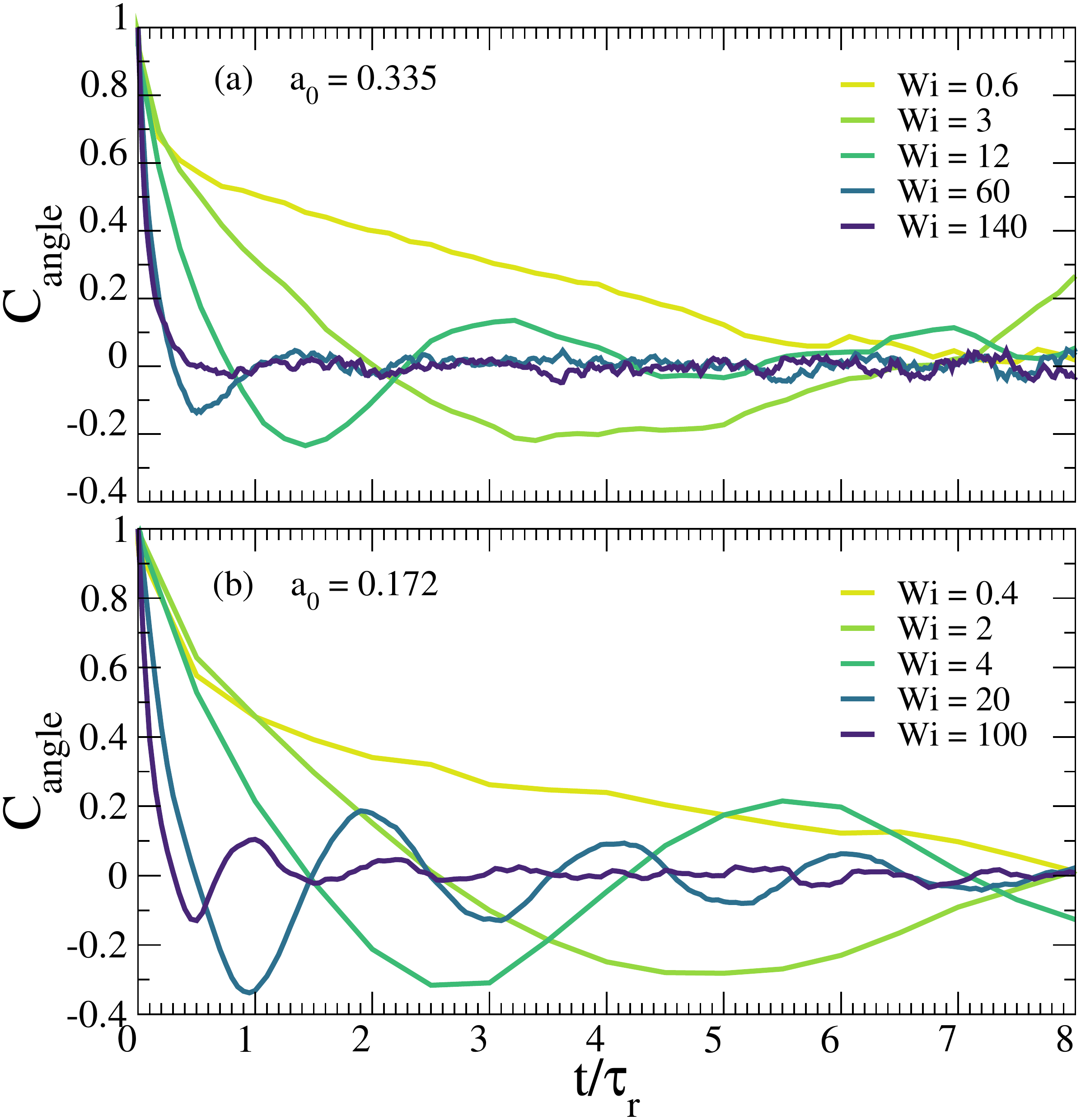}
  \caption{Angular auto-correlation function $C_{\rm angle}(t)$ for a SCNP of (a) intermediate and (b) low equilibrium asphericity.
  Times are rescaled by the longest relaxation time of the SCNP, $\tau_{\rm r}$.}
  \label{fgr:Cangle}
\end{figure}

We calculate both the flow-gradient extensional cross-correlation function $C_{xy}(t)$ as well as the angular auto-correlation function $C_{\rm angle}(t)$ for different SCNP topologies.  \ref{fgr:Cxy,fgr:Cangle} show $C_{xy}(t)$ and $C_{\rm angle}(t)$, respectively, for two SCNPs of intermediate (panels (a)) and  low equilibrium asphericity (panels (b)). We find that both SCNPs show signs of a coexistence of tumbling and tank-treading behaviors
at low and intermediate $Wi$. It should be noted that the periods of these motions differ with respect to the specific relaxation times $\tau_{\rm r}$ of the polymers (see data in \ref{fgr:Cxy,fgr:Cangle} where times are normalized by $\tau_{\rm r}$). Furthermore, the behavior of both SCNPs differs at high Weissenberg numbers.
The SCNP of intermediate $a_0$ stops tank-treading and instead just performs tumbling cycles. The SCNP of low $a_0$ continues tank-treading up to the highest shear rate simulated (corresponding to $Wi = 100$ for this SCNP), while the tumbling anti-correlation peak is lost already at $Wi = 50$ and instead shows a small positive peak centered around $t = 0$, suggesting some degree of simultaneity in the expansions (or contractions) in the gradient and flow direction. We note that SCNPs of higher equilibrium asphericity exhibit essentially the same behavior as the ones of intermediate asphericity, but the loss of tank-treading signatures in $C_{\rm angle}$ (not shown) occurs at a lower $Wi$.

\begin{figure}[ht]
\centering
  \includegraphics[width=0.5\linewidth]{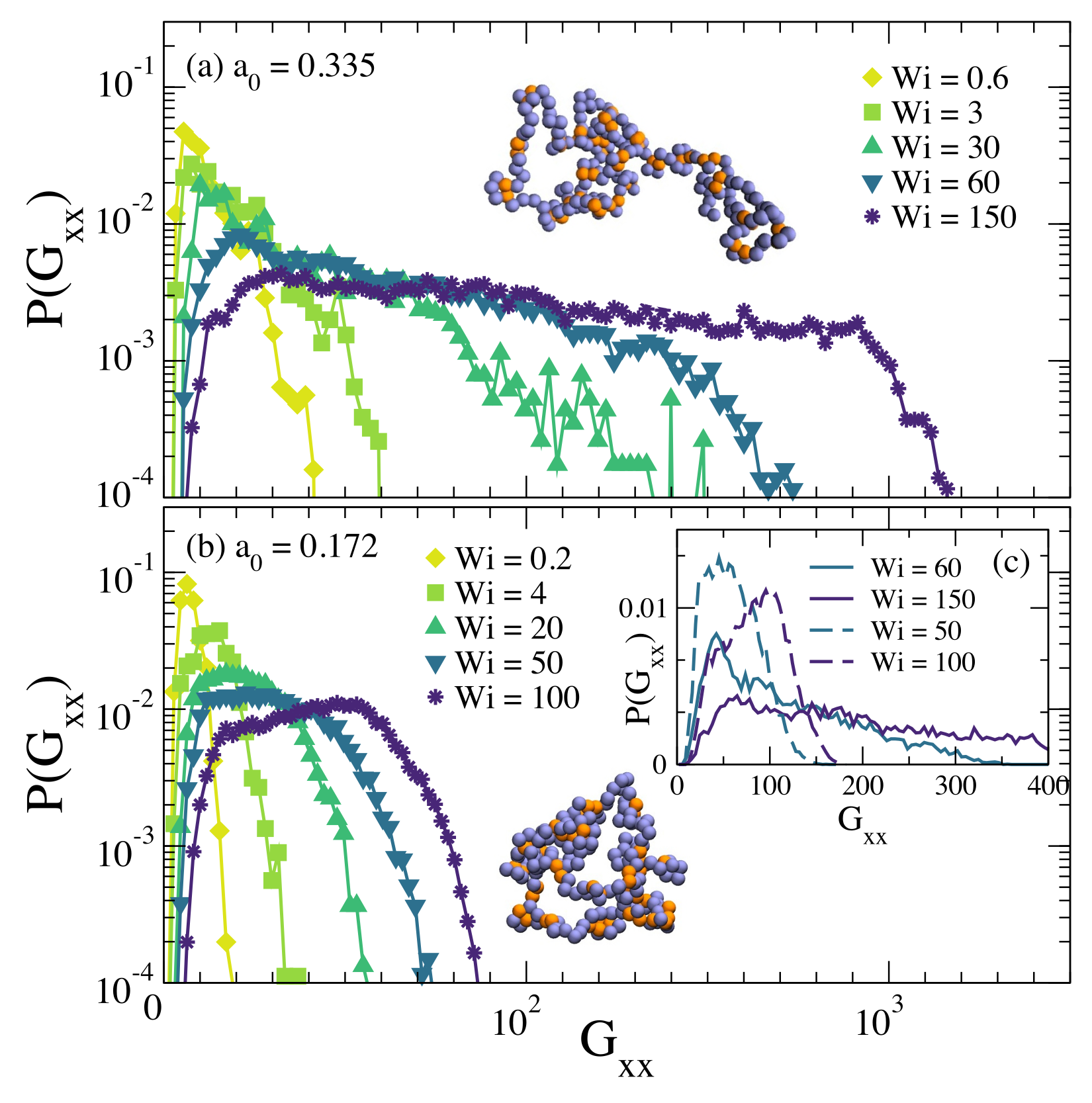}
  \caption{Probability distribution of the diagonal component  of the gyration tensor in the flow direction, $G_{xx}$, as a function of the Weissenberg number $Wi$ for a SCNP of (a) intermediate and (b) low equilibrium asphericity (log-lin scale). The inset (c) compares $P(G_{xx})$ of the two topologies at intermediate and high Weissenberg numbers on a lin-lin scale. Full lines in this inset represent the SCNP of intermediate asphericity ($a_0 = 0.335$), while dashed lines represent the SCNP of low asphericity ($a_0 = 0.172$). Snapshots of typical equilibrium conformations are included to highlight the different topologies. }
  \label{fgr:Gxxhist}
\end{figure}

To further investigate the difference in the dynamics between these two distinct SCNP topologies, we also plot (\ref{fgr:Gxxhist}) the distribution of the diagonal component of the gyration tensor in the flow direction, $P(G_{xx})$. We find that the intermediate asphericity SCNP (panel (a)) exhibits a significantly broader distribution and a less pronounced maximum across all Weissenberg numbers (note the logarithmic scale). The inset (\ref{fgr:Gxxhist}c) combines the data for $P(G_{xx})$ for both SCNPs in one graphic, but only for the two highest shear rates considered and in linear scales to highlight the maxima of the distributions. For the intermediate $a_0$ topology the maximum of the distribution does not shift much upon increasing the shear rate, but the tail of the distribution extends further to high $G_{xx}$ values. The opposite is true for the low $a_0$ SCNP: by increasing $Wi$ the width of the distribution $P(G_{xx})$ broadens only slightly, while the maximum shifts towards greater extension in the flow direction. These results are consistent with the interpretation that at high shear rates tumbling dominates the dynamics of the intermediate asphericity SCNP, whereas tank-treading is predominant in the low asphericity SCNP. Movies are included in the Supp. Info. (see description there) for the two former SCNPs, to illustrate the characteristic tumbling and tank-treading motions at high $Wi$. An example is also included for the low asphericity case
at an intermediate $Wi$,  for which tumbling (vanishing at higher $Wi$) still contributes significantly.

\begin{figure}[ht]
\centering
  \includegraphics[width=0.5\linewidth]{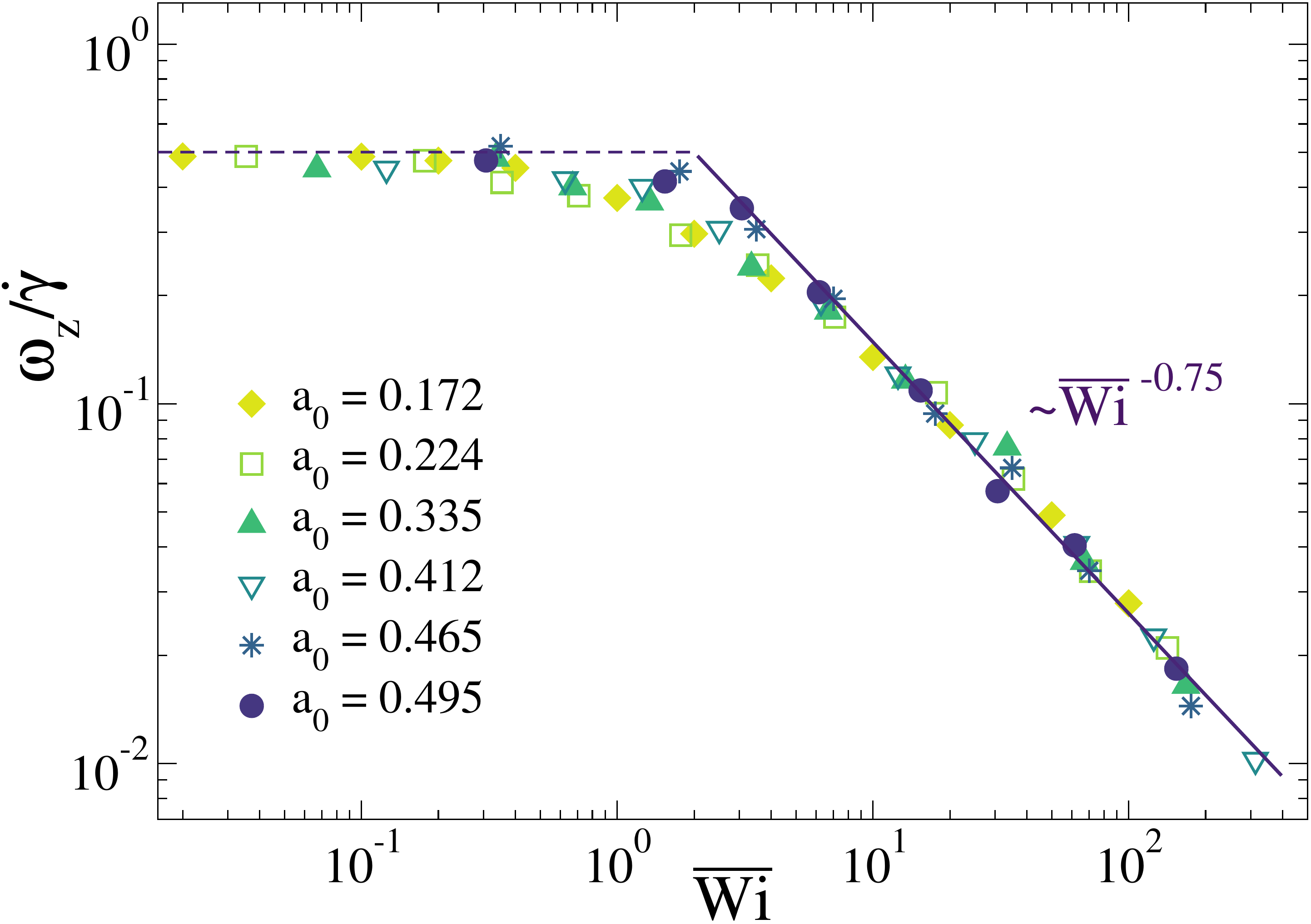}
  \caption{Rotational frequency $\omega_z$ as a function of the rescaled Weissenberg number $\overline{Wi}$
  for SCNPs with different equilibrium asphericities.}
  \label{fgr:wg}
\end{figure}

While the flow-gradient extensional cross-correlation function $C_{xy}(t)$ is used extensively in the literature to define tumbling behavior and has been linked to it through imaging studies of DNA\cite{Teixeira2005}, it does not in itself give any information about rotational dynamics. This has been pointed out by Sabli\'c \textit{et al.}\cite{Sablic2017a} in a study of the dynamical behavior of star polymers, in which they relate the anti-correlation peaks in $C_{xy}(t)$ with `breathing' modes of the star rather than tumbling. In this regard, we calculate the rotational frequency of the SCNPs in the vorticity direction by making use of the geometrical approximation \cite{Aust2002, Singh2012} 
\begin{equation}
\omega_z / \dot{\gamma} = \frac{\langle G_{yy}\rangle}{\langle G_{xx}\rangle + \langle G_{yy}\rangle} \, , 
\label{eq:rotfrequency}
\end{equation}
which is based on the rigid-body relation $\vec{\bf L} = {\bf J}\vec{\bf \omega_L}$ and relates the angular momentum $\vec{\bf L}$ to the rotational frequency $\vec{\bf \omega_L}$ \textit{via } the inertia tensor ${\bf J}$. In the derivation of the geometrical approximation  of \ref{eq:rotfrequency} it is assumed that the velocity of the monomers is  governed purely by the undisturbed velocity profile of the fluid ($v_x \simeq \dot{\gamma}y$) and that rotation only manifests around the vorticity axis, i.e. $\omega_x = \omega_y \approx 0$.  Sabli\'c \textit{et al.}\cite{Sablic2017a} compiled data from studies of different polymer architectures, bonding potentials as well as hydrodynamic simulation techniques, and found good agreement between $\omega_L$ and $\omega_z$ over a broad range of Weissenberg numbers.

Since the rotational frequency of soft objects is expected to scale linearly with the shear rate for low $Wi$, we report our results as a reduced rotational frequency $\omega_z/\dot{\gamma}$ versus $\overline{Wi}$ in \ref{fgr:wg}. The data are found to approach the expected 
linear scaling $\omega_z \simeq \dot{\gamma}/2$ at small $\overline{Wi}$ for all SCNP topologies. For large shear rates, the data collapse onto a master curve as well, when plotted against the rescaled Weissenberg number $\overline{Wi}$, and scale as $\omega_z/\dot{\gamma} \sim Wi^{-0.75}$. This common scaling for the rotational frequency is somewhat unexpected, 
given the rather different dynamic behaviors displayed at large $Wi$ by the $C_{xy}(t)$ and $C_{\rm angle}(t)$ correlators (with predominance of tumbling or tank-treading motions depending on the SCNP topology).  
Still, one should keep in mind that both $\omega_L$ and $\omega_z$ are based on a generalization of rigid-body rotations to soft objects, and therefore correspond to the rotation of a rigid-body having the average shape of the polymer. Interpretation of $\omega_z$ as an angular velocity to quantify either the tumbling or the tank-treading frequency suffers from the fact that rotational vibrations are included in the calculation of $\omega_z$, which do not add to the molecule overall rotation. Recent studies\cite{Sablic2017b, Jaramillo-Cano2018a} have suggested to use the co-rotating Eckart frame\cite{Eckart1935} to decouple rotations from vibrations and thus better understand the dynamics of soft objects. An in-depth analysis of the rotational dynamics of the various SCNP topologies in terms of the Eckart formalism is beyond the scope of this work and will be studied in a future work.
As can bee seen in \ref{tab:scaling}, the exponents  found for the $Wi$-dependence of the rotational frequency 
(as defined in \ref{eq:rotfrequency}) for the different topologies show a strong dispersion and no obvious trend.  
The use of the Eckart frame to determine the rotational frequency might shed light on this question.

\section{IV. Conclusions}

By means of a coarse-grained polymer model, combined with multi-particle collision dynamics
to implement hydrodynamic interactions, we have investigated SCNPs under homogeneous shear flow.
SCNPs emerge as a novel class of complex macromolecular objects with a response to shear that is distinct from other polymeric objects
such as linear chains, rings, dendrimers or stars.  This is demonstrated by the unique set of scaling exponents for the shear rate dependence of 
static and dynamic observables as the components of the gyration tensor, orientational resistance, intrinsic viscosity or rotational frequency.
Surprisingly, the obtained sets of exponents are, at most, marginally dependent on the specific topology of the SCNP (globular or sparse).
This suggests that the response of SCNPs to shear is inherently related to the network-like character of their molecular architecture,
but not the specific connectivity of the network. By analyzing adapted time correlation functions we have found that at high Weissenberg numbers 
the dynamics of the sparse SCNPs is dominated by tumbling motion.  Tank-treading is predominant for the most globular SCNPs. 

The general physical scenario presented here may motivate not only experimental tests in SCNPs but also in intrinsically disordered proteins,
given the observed structural similarities between both systems \cite{Moreno2016JPCL,Moreno2018}.
Another question to be investigated is the effect of the concentration of SCNPs on their response to shear. Investigations in semidilute solutions
(up to a few times the overlap concentration) of unentangled or weakly entangled linear chains \cite{Huang2012mac,Huang2012jpcm} 
and star polymers \cite{Fedosov2012,Singh2012,Singh2013} have shown  small or moderate changes 
in the characteristic exponents for the $Wi$-dependence with respect to those found at high dilution.
It is not obvious how the high-dilution scenario for the SCNP behavior under shear flow will be affected by increasing concentration,
since in equilibrium SCNPs are indeed more strongly perturbed by crowding (collapsing to fractal globular 
structures \cite{Moreno2016JPCL,GonzalezBurgos2018}) than linear chains. Work in this direction is in progress.

\begin{acknowledgement}

We acknowledge financial support from the projects MAT2015-63704-P (MINECO-Spain and FEDER-UE) and IT-654-13 (Basque Government, Spain).


\end{acknowledgement}



\providecommand{\latin}[1]{#1}
\providecommand*\mcitethebibliography{\thebibliography}
\csname @ifundefined\endcsname{endmcitethebibliography}
  {\let\endmcitethebibliography\endthebibliography}{}

\newpage

\begin{center}
\Large
{\bf SUPPORTING INFORMATION}
\end{center}
\normalsize
\vspace{1 cm}

\noindent {\bf Static and dynamic observables as a function of the Weissenberg number, for linear chains of $N = 200$ monomers}
\vspace{1cm}
\begin{figure}[h!]
\centering
  \includegraphics[width=0.62\linewidth]{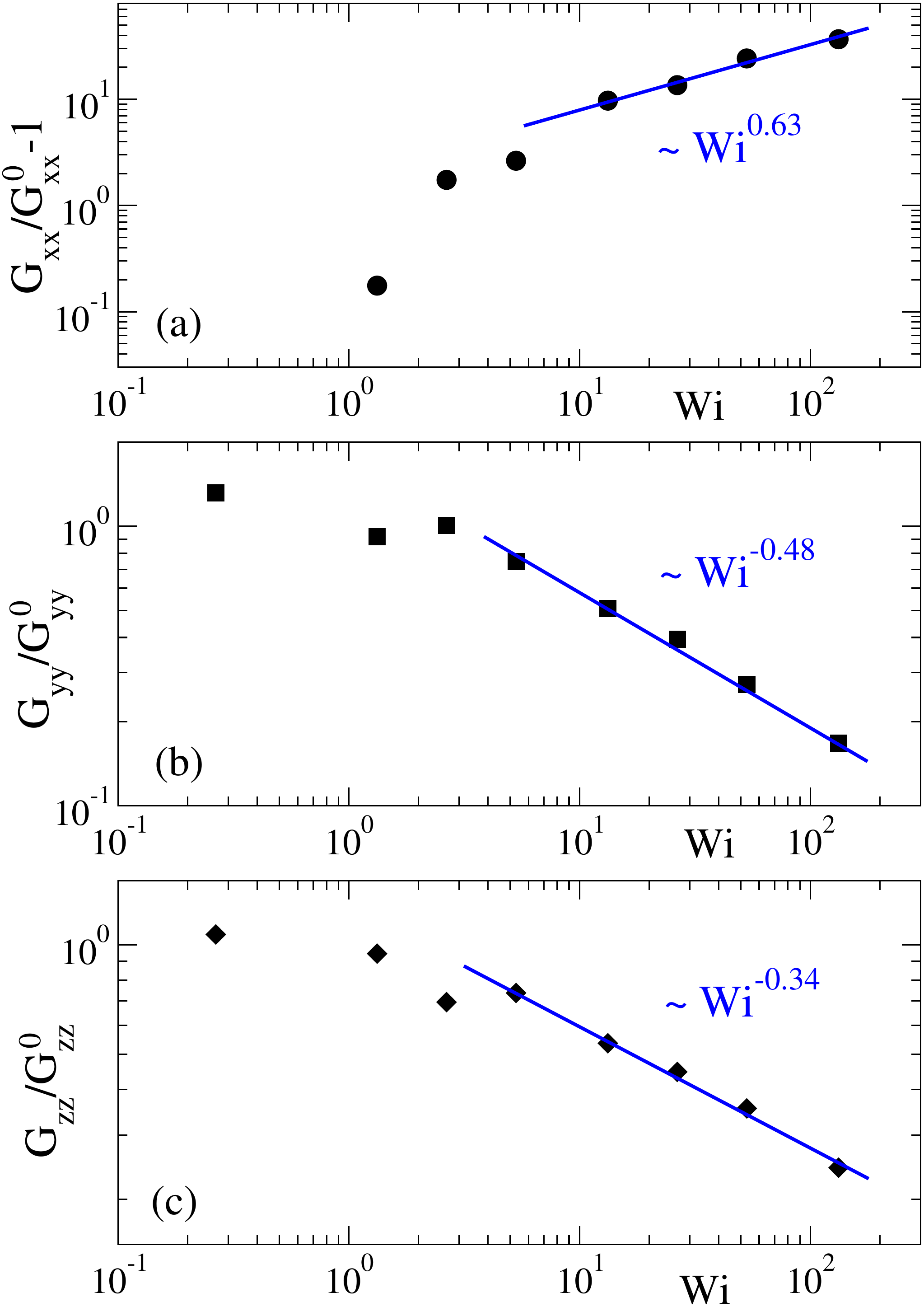}\\  
\vspace{4 mm} Figure S1. Normalized diagonal components of the gyration tensor.
\end{figure}

\newpage

\begin{figure}[h!]
\centering
  \includegraphics[width=0.57\linewidth]{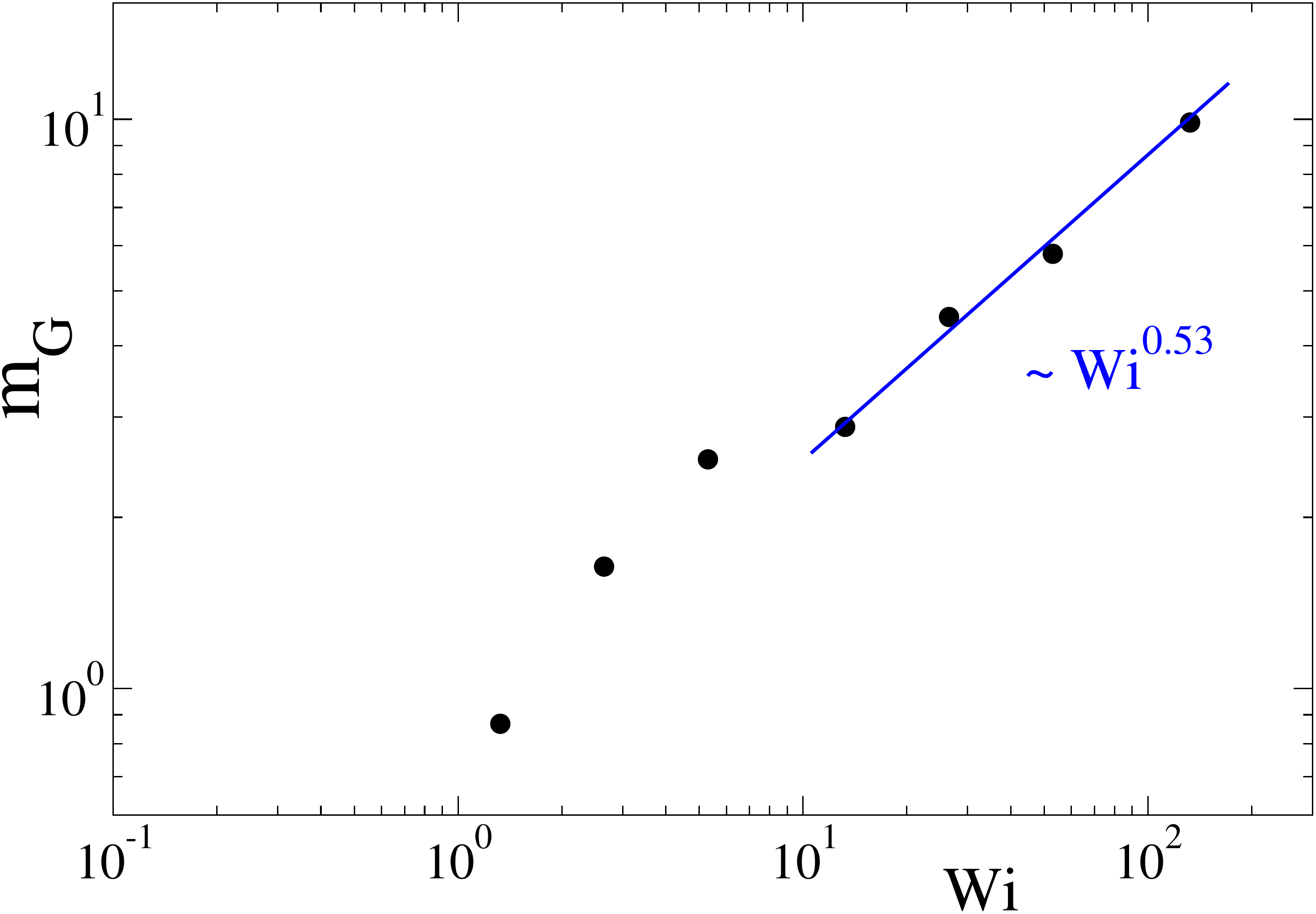} \\
\vspace{4 mm} Figure S2. Orientational resistance.
\end{figure}
\vspace{3 mm}

\begin{figure}[h!]
\centering
  \includegraphics[width=0.50\linewidth]{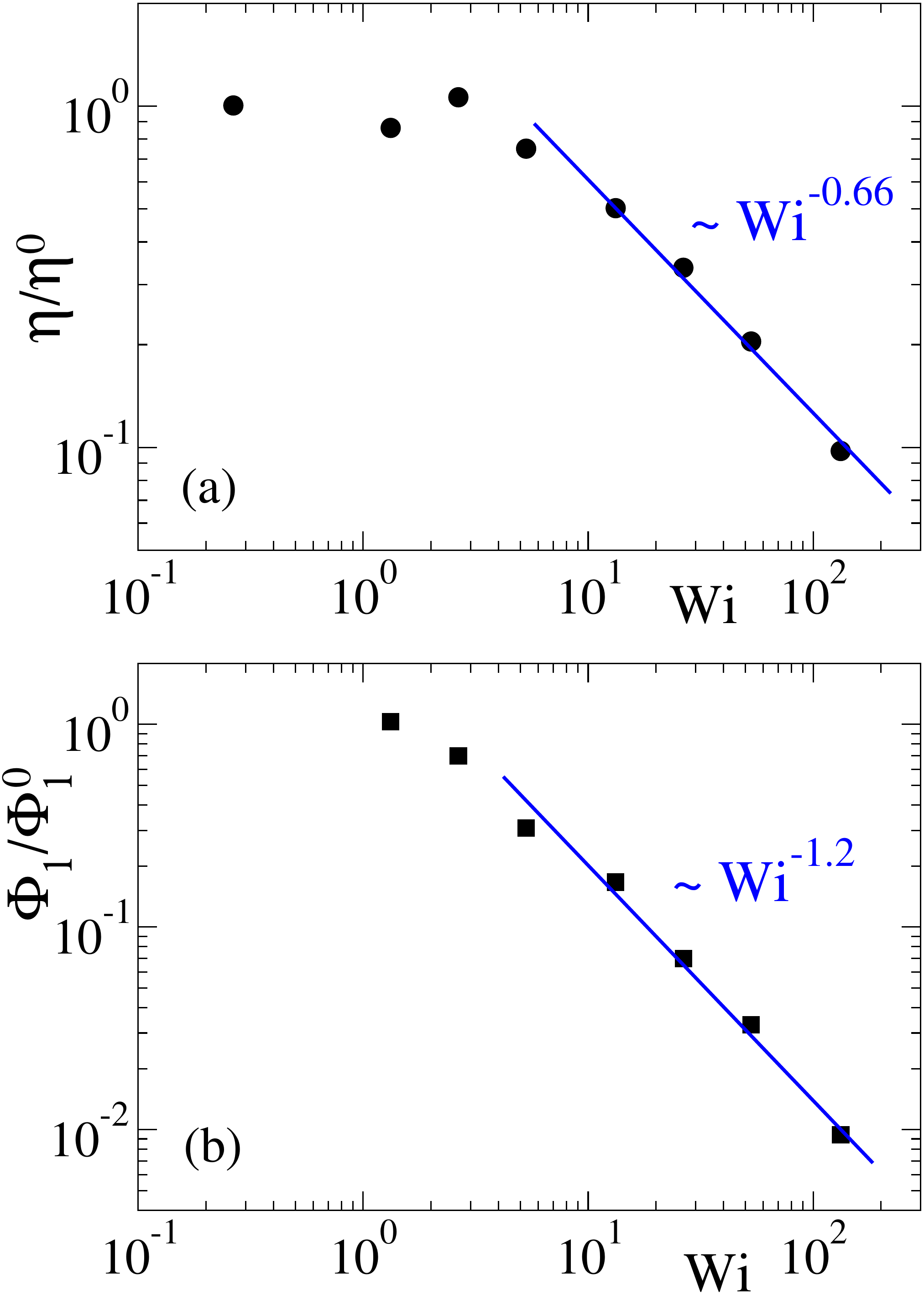} \\ 
\vspace{4 mm} Figure S3. Normalized viscosity (a) and first normal stress coefficient (b).
\end{figure}

\newpage

\begin{figure}[h!]
\centering
  \includegraphics[width=0.57\linewidth]{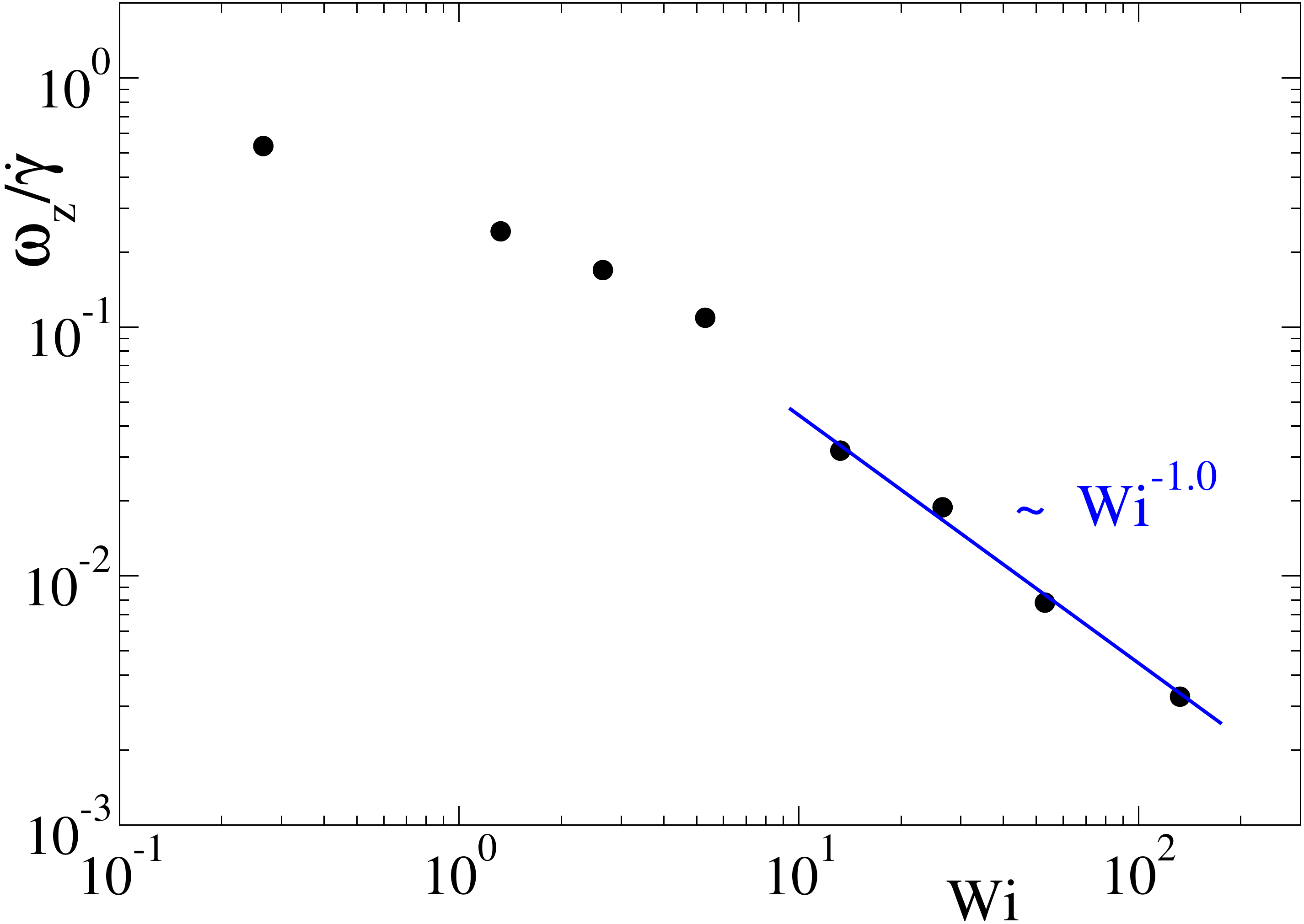} \\ 
\vspace{4 mm} Figure S4. Rotational frequency. 
\end{figure}

\end{document}